\documentclass[ pra,letter,  paper,superscriptaddress]{revtex4-2}
\usepackage{graphicx,amsmath,amssymb,amsfonts,latexsym,xcolor,dcolumn,bm}
\usepackage{appendix}
\usepackage[colorlinks]{hyperref}

\begin{document}

\global\long\def\eqn#1{\begin{align}#1\end{align}}
\global\long\def\vec#1{\overrightarrow{#1}}
\global\long\def\ket#1{\left|#1\right\rangle }
\global\long\def\bra#1{\left\langle #1\right|}
\global\long\def\bkt#1{\left(#1\right)}
\global\long\def\sbkt#1{\left[#1\right]}
\global\long\def\cbkt#1{\left\{#1\right\}}
\global\long\def\abs#1{\left\vert#1\right\vert}
\global\long\def\cev#1{\overleftarrow{#1}}
\global\long\def\der#1#2{\frac{{d}#1}{{d}#2}}
\global\long\def\pard#1#2{\frac{{\partial}#1}{{\partial}#2}}
\global\long\def\re{\mathrm{Re}}
\global\long\def\im{\mathrm{Im}}
\global\long\def\dd{\mathrm{d}}
\global\long\def\ddd{\mathcal{D}}
\global\long\def\avg#1{\left\langle #1 \right\rangle}
\global\long\def\mr#1{\mathrm{#1}}
\global\long\def\mb#1{{\mathbf #1}}
\global\long\def\mc#1{\mathcal{#1}}
\global\long\def\tr{\mathrm{Tr}}
\global\long\def\dbar#1{\stackrel{\leftrightarrow}{\mathbf{#1}}}

\global\long\def\nth{$n^{\mathrm{th}}$\,}
\global\long\def\mth{$m^{\mathrm{th}}$\,}
\global\long\def\non{\nonumber}

\newcommand{\orange}[1]{{\color{orange} {#1}}}
\newcommand{\cyan}[1]{{\color{cyan} {#1}}}

\newcommand{\blue}[1]{{\color{blue} {#1}}}
\newcommand{\yellow}[1]{{\color{yellow} {#1}}}
\newcommand{\green}[1]{{\color{green} {#1}}}
\newcommand{\red}[1]{{\color{red} {#1}}}

\newcommand{\teal}[1]{{\color{teal} {#1}}}
\newcommand{\ks}[1]{{\textcolor{teal}{[KS: #1]}}}

\newcommand{\bd}{{\bf d}}      
\newcommand{\bv}{{\bf v}}
\newcommand{\hbp}{\hat{\bp}}
\newcommand{\hbx}{\hat{\bx}}
\newcommand{\hq}{\hat{q}}
\newcommand{\hp}{\hat{p}}
\newcommand{\ha}{\hat{a}}
\newcommand{\had}{{a}^{\dag}}
\newcommand{\ad}{a^{\dag}}
\newcommand{\hsig}{{\hat{\sigma}}}
\newcommand{\nt}{\tilde{n}}
\newcommand{\itf}{\sl}
\newcommand{\eps}{\epsilon}
\newcommand{\bsig}{\pmb{$\sigma$}}
\newcommand{\beps}{\pmb{$\eps$}}
\newcommand{\bmu}{\pmb{$ u$}}
\newcommand{\balpha}{\pmb{$\alpha$}}
\newcommand{\bbeta}{\pmb{$\beta$}}
\newcommand{\bgamma}{\pmb{$\gamma$}}
\newcommand{\bu}{{\bf u}}
\newcommand{\bpi}{\pmb{$\pi$}}
\newcommand{\bSig}{\pmb{$\Sigma$}}
\newcommand{\be}{\begin{equation}}
\newcommand{\ee}{\end{equation}}
\newcommand{\bea}{\begin{eqnarray}}
\newcommand{\eea}{\end{eqnarray}}
\newcommand{\sss}{_{{\bf k}\lambda}}
\newcommand{\ssss}{_{{\bf k}\lambda,s}}
\newcommand{\dip}{\langle\sigma(t)\rangle}
\newcommand{\dipp}{\langle\sigma^{\dag}(t)\rangle}
\newcommand{\sig}{{\tilde{\sigma}}}
\newcommand{\sigd}{{\sigma}^{\dag}}
\newcommand{\sigz}{{\sigma_z}}
\newcommand{\ra}{\rangle}
\newcommand{\la}{\langle}
\newcommand{\om}{\omega}
\newcommand{\Om}{\Omega}
\newcommand{\pa}{\partial}
\newcommand{\bR}{{\bf R}}
\newcommand{\bx}{{\bf x}}
\newcommand{\br}{{\bf r}}
\newcommand{\bE}{{\bf E}}
\newcommand{\bH}{{\bf H}}
\newcommand{\bB}{{\bf B}}
\newcommand{\bP}{{\bf P}}
\newcommand{\bD}{{\bf D}}
\newcommand{\bA}{{\bf A}}
\newcommand{\bek}{{\bf e}\rmk}
\newcommand{\rmk}{_{{\bf k}\lambda}}
\newcommand{\rk}{_{{\bf k}_1{\lambda_1}}}
\newcommand{\rkk}{_{{\bf k}_2{\lambda_2}}}
\newcommand{\rkz}{_{{\bf k}_1{\lambda_1}z}}
\newcommand{\rkkz}{_{{\bf k}_2{\lambda_2}z}}
\newcommand{\bsij}{{\bf s}_{ij}}
\newcommand{\bk}{{\bf k}}
\newcommand{\bp}{{\bf p}}
\newcommand{\epso}{{1\over 4\pi\eps_0}}
\newcommand{\BB}{{\mathcal B}}
\newcommand{\AAA}{{\mathcal A}}
\newcommand{\NN}{{\mathcal N}}
\newcommand{\mm}{{\mathcal M}}
\newcommand{\RR}{{\mathcal R}}
\newcommand{\bS}{{\bf S}}
\newcommand{\bL}{{\bf L}}
\newcommand{\bJ}{{\bf J}}
\newcommand{\bI}{{\bf I}}
\newcommand{\bF}{{\bf F}}
\newcommand{\bsub}{\begin{subequations}}
\newcommand{\esub}{\end{subequations}}
\newcommand{\baline}{\begin{eqalignno}}
\newcommand{\ealine}{\end{eqalignno}}
\newcommand{\isat}{{I_{\rm sat}}}
\newcommand{\Is}{I^{\rm sat}}
\newcommand{\Ip}{I^{(+)}}
\newcommand{\Imm}{I^{(-)}}
\newcommand{\Inu}{I_{\nu}}
\newcommand{\bInu}{\overline{I}_{\nu}}
\newcommand{\bN}{\overline{N}}
\newcommand{\qnu}{q_{\nu}}
\newcommand{\oqn}{\overline{q}_{\nu}}
\newcommand{\qsat}{q^{\rm sat}}
\newcommand{\Iout}{I_{\nu}^{\rm out}}
\newcommand{\topt}{t_{\rm opt}}
\newcommand{\crr}{{\mathcal{R}}}
\newcommand{\cE}{{\mathcal{E}}}
\newcommand{\cH}{{\mathcal{H}}}
\newcommand{\epsoo}{\epsilon_0}
\newcommand{\ombar}{\overline{\om}}
\newcommand{\cEp}{{\mathcal{E}}^{(+)}}
\newcommand{\cEm}{{\mathcal{E}}^{(-)}}
\newcommand{\bvv}{\tilde{\bv}}
\newcommand{\pr}{^{\prime}}
\newcommand{\dpr}{^{\prime\prime}}
\newcommand{\hk}{\hat{\bk}}
\newcommand{\hn}{\hat{\bf n}}
\newcommand{\ok}{\om_1}
\newcommand{\okk}{\om_2}

\title{Dipoles in blackbody radiation: Momentum fluctuations, decoherence, and drag force}
\author{Kanupriya Sinha}
\affiliation{School of Electrical, Energy and Computer Engineering, Arizona State University, Tempe, Arizona 85281 USA}
\email{kanu.sinha@asu.edu}
\author{Peter W. Milonni}
\affiliation{Theoretical Division (T-4), Los Alamos National Laboratory, Los Alamos, New Mexico 87545 USA} 
\affiliation{Department of Physics and Astronomy, University of Rochester, Rochester, NY 14627 USA}
\email{peter\_milonni@comcast.net} 

\begin{abstract}
A general expression is derived for the momentum diffusion constant of a small polarizable particle in blackbody radiation, and is shown to be closely related to the long-wavelength collisional decoherence rate for such a particle in a thermal environment. We show how this diffusion constant appears in the steady-state photon emission rate of two dipoles induced by blackbody radiation. We consider in addition the Einstein--Hopf drag force on a small polarizable particle moving in a blackbody field, and derive its fully relativistic form from the Lorentz transformation of forces.

\end{abstract} 
\maketitle
\section{Introduction}

One of the ways in which the quantized nature of the electromagnetic field is manifested is in the recoil of particles in scattering, absorption, and emission processes \cite{shore}. In spontaneous emission, for example, a classical treatment of the field would not account for the recoil of an atom because the field from the atom would have zero linear momentum, a consequence of the inversion symmetry of the field about the atom. Einstein \cite{ein1} showed that atoms in an isotropic field with spectral energy density $\rho(\om)$ undergo a mean-square momentum gain resulting from the recoil momentum of magnitude $\hbar\om/c$ in each emission and absorption event. Assuming the atoms have a mean kinetic energy $\avg{\frac{1}{2}m{\bf v}^2}=\frac{3}{2}k_BT$ in equilibrium, this mean-square momentum increase is compensated by a drag force in just such a way as to yield the Planck distribution for $\rho(\om)$. In earlier work, Einstein and Hopf \cite{eh1,eh2} obtained classical expressions for the variance of the fluctuating momentum and the drag force on a small polarizable particle in blackbody radiation and used these expressions to derive the Rayleigh--Jeans spectrum. In this paper we derive a simple quantum-mechanical generalization of their expression for the momentum variance and diffusion constant. A similar generalization of the Einstein--Hopf formula for the drag force was obtained earlier \cite{mkr}. We show that, not surprisingly, these two generalizations of Einstein's expressions can be used to extend Einstein's analysis for atoms \cite{ein1} to any small polarizable particle in thermal equilibrium with radiation. 

 The expression we obtain in the Heisenberg picure for the momentum diffusion constant is an ensemble average over random momentum kicks. We show that this momentum diffusion constant appears also in the rate of photon emission from two localized dipoles in blackbody radiation, and has the same form as the center-of-mass decoherence rate of a particle in a thermal photon environment. Such blackbody-induced decoherence was previously analyzed in a scattering-theory framework \cite{joos}, wherein which-path information carried by the thermal photons scattering off the particle causes the particle's center of mass to decohere and localize in the position basis. If the characteristic thermal wavelength $\lambda_\mr{th}$ of the electromagnetic environment is larger than the coherence length $\Delta x$, each  scattered thermal photon only gains partial information about the particle's position (`long-wavelength limit'); for shorter wavelengths a single scatterer has sufficient information to resolve a coherent superposition (`short-wavelength limit'). It will be shown that the momentum diffusion constant is related specifically to the long-wavelength limit of the blackbody-induced center-of-mass decoherence rate.

 The rest of the paper is organized as follows. In Section \ref{sec:momentum} we calculate the momentum variance and diffusion constant based first on a model in which the particle experiences random, instantaneous, and statistically independent momentum kicks. This is followed by a more rigorous calculation which is simplified by treating the blackbody field at first as a classical stochastic field as in the original Einstein--Hopf theory \cite{eh1,eh2}. Following Einstein and Hopf we treat the electric field and its spatial derivative as independent stochastic processes; a proof of this independence is given in Appendix \ref{App:stind}. In Section \ref{sec:decoh} we calculate the photon emission rate for two dipoles in blackbody radiation. We show how the reduction with the dipole separation of an interference term in this rate involves the diffusion constant.  In Section \ref{sec:drag} we derive the relativistic form of the drag force on a moving particle in blackbody radiation directly from the Lorentz transformation of forces. We discuss in Section \ref{sec:discussion} some aspects of our calculations as they  relate to previous work, and conclude with remarks relating to Einstein's derivation of the Planck spectrum and the applicability of our results to atoms in blackbody radiation.   Appendix~\ref{App:air} gives a brief derivation of the  momentum diffusion constant due to scattering of air molecules and the corresponding  center-of-mass decoherence rate.

 \section{\label{sec:momentum} Momentum Fluctuations}

\subsection{\label{sec:kicks} Momentum Kicks}

\begin{figure}[t]
    \centering
    \includegraphics[width = 2.25 in]{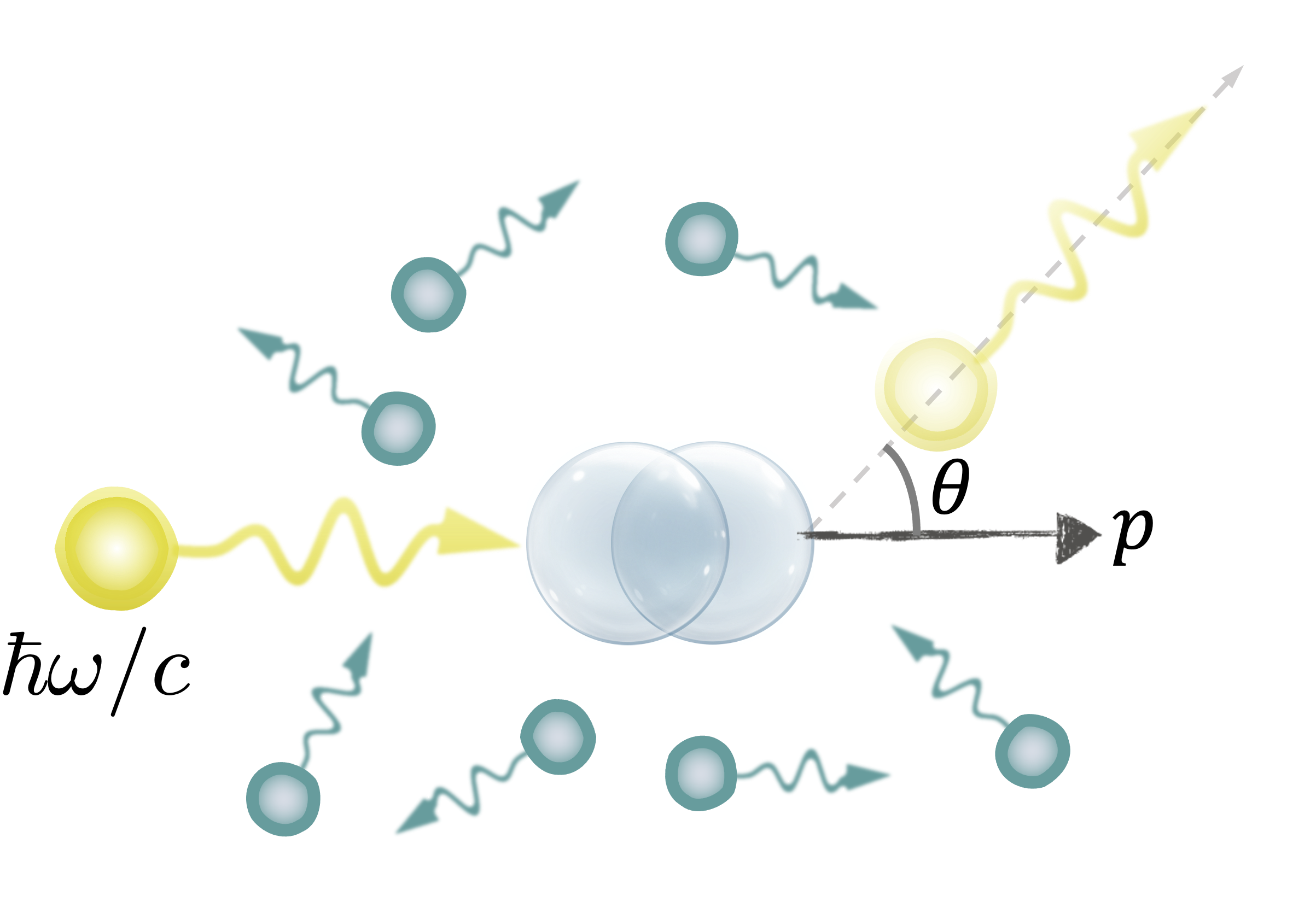}
    \caption{Schematic representation of a polarizable particle interacting with a thermal electromagnetic field.}
    \label{fig:Sch}
\end{figure}

We begin by assuming that the momentum diffusion of a particle results from statistically independent momentum kicks of magnitude $\hbar\om/c$ due to photons of frequency $\om$ (see Fig.~\ref{fig:Sch}). Together with a frictional force $-\beta p(t)$, this implies, for the cumulative change in the linear momentum $p(t)$,
\be
\dot{p}(t)+\beta p(t)=\sum_{t_j}\frac{\hbar\om}{c}\delta(t-t_j),
\label{eq201}
\ee
for motion along, say, the $x$ direction, where $t_j$ is the (random) time at which the $j$th instantaneous momentum kick occurs. Thus in the steady-state limit
\be
p(t)=\sum_{t_j}\frac{\hbar\om}{c}G(t-t_j),
\label{eq202}
\ee 
where $G(t) \equiv e^{- \beta t}$ is the response to a unit impulse.

The field energy density in the interval $[\om,\om+d\om]$ is given by $\rho(\om)d\om$. The contribution of photons with energy between $[\om,\om+d\om]$ to the steady-state, mean-square momentum of the particle is given, according to Campbell's theorem \cite{campbell,rice}, by
\be
d\la \Delta p^2\ra=\bkt{\frac{\hbar\om}{c}}^2\times {\rm{(average \ photon \ rate}} \ u(\om)d\om) \times \int_{0}^{\infty}G^2(t)dt
\label{eq203}.
\ee
For $u(\om)d\om$ we use the photon rate per unit area per unit time, $c\rho(\om)d\om/\hbar\om$, times the Rayleigh cross section 
\be
\sigma(\om)=\frac{8\pi}{3}\bkt{\frac{\om}{c}}^4\abs{\alpha(\om)}^2=\frac{4\pi\om}{c}\alpha_I(\om),
\label{eq204}
\ee
where $\alpha(\om)$ is the particle's polarizability \cite{scalar}, $\alpha_I(\om)$ is its imaginary part,  and we have used the optical theorem for Rayleigh scattering \cite{QObook}:
\be
\alpha_I(\om)=\frac{2}{3}\bkt{\frac{\om}{c}}^3|\alpha(\om)|^2.
\label{eq206}
\ee
 We integrate Eq.~\eqref{eq203} over all field frequencies, using the relation \cite{QObook} \eqn{\rho(\om)=\frac{\hbar\om^3}{\pi^2c^3}n(\om)
\label{pla}
}
between $\rho(\om)$ and the average number $n(\om)$ of photons at frequency $\om$, to obtain

\be
\la \Delta p^2\ra=\frac{4\hbar^2}{\pi c^5}\int_0^{\infty}d\om\om^5\alpha_I(\om)n(\om)\times 1/ (2\beta).
\label{eq208}
\ee
$2\beta$ is the rate at which $\la \Delta p^2\ra$ decreases due to the drag force, and in steady state the rate $\la \Delta p^2\ra/\Delta t$ at which $\la\Delta p^2\ra$
increases must equal $2\beta\la \Delta p^2\ra$. It follows that
\be
\frac{\la \Delta p^2\ra}{\Delta t}=\frac{4\hbar^2}{\pi c^5}\int_0^{\infty}d\om\om^5\alpha_I(\om)n(\om).
\label{eq209}
\ee

We have assumed momentum kicks $\hbar\om/c$, but for any particular kick the recoil momentum depends on the photon scattering angle $\theta$. Let $\hat{k}$ and $\hat{k}^{\prime}$ be the unit vectors for the directions of incoming and scattered photons, respectively. The recoil momentum of the particle in the $\hat{x}$ direction for this scattering event is
\be
\delta p=\frac{\hbar\om}{c}\bkt{\hat{k}-\hat{k}^{\prime}}\cdot
\hat{x}.
\label{reco1}
\ee
In terms of the scattering angle $\theta$, $|\hat{k}-\hat{k}^{\prime}|^2=4\sin^2(\theta/2)$ and
\be
\overline{\delta p^2}=\frac{1}{3}\bkt{\frac{\hbar\om}{c}}^2
\bkt{4\sin^2\bkt{\theta/2}}
\ee
when we average over the directions of $\hat{k}-\hat{k}^{\prime}$. The average of this expression over all scattering angles is then $(2/3)(\hbar\om/c)^2$. We therefore account for all possible scattering angles by multiplying the expression (\ref{eq209}) by 2/3:
\be
\frac{\la\Delta p^2\ra}{\Delta t}=\frac{8\hbar^2}{3\pi c^5}\int_0^{\infty}d\om\om^5\alpha_I(\om)n(\om).
\label{eq210}
\ee

This derivation of the momentum diffusion constant ${\la \Delta p^2\ra}/{\Delta t}$ based on Campbell's theorem is simpler and more intuitive than the more rigorous derivation given below. However, it does not account for the Bose--Einstein statistics of blackbody photons, the effect of which is to replace $n(\om)$ in Eq.~\eqref{eq210} by $n(\om)[n(\om)+1]$. The application of Campbell's theorem  requires that successive impulses are statistically independent, whereas the ``photon bunching" effect in blackbody radiation implies that successive momentum kicks are not in fact independent. The effect of the Bose--Einstein factor $n(\om)+1$ is considered for two examples in the following section. For these examples the effect of the Bose--Einstein factor is negligible and Campbell's theorem provides an accurate estimate of the momentum diffusion constant.

It might be worth noting that while Campbell's theorem does not account for Bose-Einstein statistics, it can be used, as shown in Appendix~\ref{App:air}, to describe momentum diffusion from other environmental scatterers such as air molecules. 

\subsection{\label{sec:induced} Calculation of the Momentum Diffusion Constant from the Force on an Induced Dipole} 
We start from the classical expression for the force on an electric dipole moment $\bd$ in an electromagnetic field: 
\be
\bF=(\bd\cdot\nabla)\bE+\frac{1}{c}\frac{\pa \bd}{\pa t}\times\bB.
\label{eq101]}
\ee 
For motion along the $x$ axis and a dipole moment $\bd=d_0\hat{z}$,
\be
F_x=d_0\frac{\pa E_x}{\pa z}-\frac{1}{c}\frac{\pa d_0}{\pa t}B_y,
\label{eq102}
\ee
and the impulse experienced by the dipole in a time interval from $t=0$ to $t=\Delta t$ is
\bea
\Delta p&=&\int_0^{\Delta t}dtd_0\frac{\pa E_x}{\pa z}-\frac{1}{c}\int_0^{\Delta t}dt\frac{\pa d_0}{\pa t}B_y
=\int_0^{\Delta t}dtd_0\frac{\pa E_x}{\pa z}-\frac{1}{c}\int_0^{\Delta t}dt\Big[\frac{\pa}{\pa t}(B_yd_0)-d_0\frac{\pa B_y}{\pa t}\Big]\nonumber\\
&=&\int_0^{\Delta t}dtd_0\frac{\pa E_z}{\pa x}-\frac{1}{c}(B_yd_0)\Big|_0^{\Delta t},
\label{eq103}
\eea
where we have used the Maxwell equation $(1/c)\pa B_y/\pa t=\pa E_z/\pa x-\pa E_x/\pa z$. The last term vanishes in a steady state, in which case
\be
\Delta p=\int_0^{\Delta t}dtd_0\frac{\pa E_z}{\pa x}.
\label{eq1033}
\ee

For the blackbody field at the point dipole we write
\be
E_z(t)=i\sum\rmk\Big(\frac{2\pi\hbar\om}{V}\Big)^{1/2}\big[\tilde{a}\rmk e^{-i\om t}-\tilde{a}^*_{\rmk}e^{i\om t}\big]e_{\bk\lambda z}
\label{eq104}
\ee
in the conventional notation \cite{notation}, except that in the classical approach we have taken thus far we regard $\tilde{a}\rmk$ and $\tilde{a}^*\rmk$ as 
{\sl classical stochastic variables} rather than photon annihilation and creation operators. Similarly,
\be 
\frac{\pa E_z}{\pa x}=-\sum\rmk\Big(\frac{2\pi\hbar\om }{V}\Big)^{1/2}k_x\big[\tilde{a}\rmk e^{-i\om t}+\tilde{a}^*_{\rmk}e^{i\om t}\big]e_{\bk\lambda z}.
\label{eq105}
\ee
Introducing the complex polarizability $\alpha(\om)$ for the induced dipole along the direction $E_z$ of the electric field, we have 
\be
d_0=i\sum\rmk\bkt{\frac{2\pi\hbar\om }{V}}^{1/2}\sbkt{\alpha(\om )\tilde{a}\rmk e^{-i\om t}-\alpha^*(\om )\tilde{a}^*_{\rmk}e^{i\om t}}e_{\bk\lambda z},
\label{eq106}
\ee
and therefore we obtain, after doing the integral over time in Eq. (\ref{eq1033}),
\bea
\Delta p&=&-2i\sum\rk\sum\rkk\bkt{\frac{2\pi\hbar\ok}{V}}^{1/2}\bkt{\frac{2\pi\hbar\okk}{V}}^{1/2}
{k_1}_x\frac{\sin\bkt{\frac{1}{2}(\ok-\okk)\Delta t}}{\ok-\okk}\nonumber\\
&&\mbox{}\times\sbkt{\alpha(\okk)\tilde{a}\rkk \tilde{a}^*\rk e^{i(\ok-\okk)\Delta t/2}-\alpha^*(\okk)\tilde{a}^*\rkk \tilde{a}\rk  e^{-i(\ok-\okk)\Delta t/2}}
{e\rkz}{e\rkkz},\nonumber\\
\label{eq107}
\eea
where we neglect non-resonant terms varying as $1/(\ok+\okk)$. The terms denoted by subscripts 1 and 2 in Eq.~\eqref{eq107} derive from  $\pa E_z/\pa x$ and $E_z$, respectively. Einstein and Hopf  \cite{eh1} showed that these terms are statistically independent.   We give a simplified proof of this independence, which significantly simplifies the calculation of $\la\Delta p^2\ra$, in Appendix~\ref{App:stind}. From this independence it follows from Eq.~\eqref{eq107} that
\bea
\la\Delta p^2\ra&=&4\sum\rk\sum\rkk\frac{2\pi\hbar\ok}{V}\frac{2\pi\hbar\okk}{V}k^2_{1x}\frac{\sin^2\bkt{\frac{1}{2}(\ok-\okk)\Delta t}}{(\ok-\okk)^2}\abs{\alpha(\okk)}^2\times
\nonumber\\
&&\mbox{}\Big[\la \tilde{a}\rkk \tilde{a}^*\rkk\ra\la \tilde{a}^*\rk\tilde{a}\rk\ra +\la\tilde{a}^*\rkk \tilde{a}\rkk\ra\la \tilde{a}\rk \tilde{a}^*\rk\ra \Big]
{e^2\rkz}{e^2\rkkz}\nonumber\\
&&\mbox{}\hspace{-1.4cm}=8\sum\rk\sum\rkk\frac{2\pi\hbar\ok}{V}\frac{2\pi\hbar\okk}{V}k^2_{1x}\frac{\sin^2\bkt{\frac{1}{2}(\ok-\okk)\Delta t}}{(\ok-\okk)^2}  A(\ok)A(\okk)\abs{\alpha(\okk)}^2 
{e^2\rkz}{e^2\rkkz},\nonumber\\
\label{eq108}
\eea
where we have assumed the ensemble averages
\bea
\la \tilde{a}\rk \tilde{a}^*_{{\bf k}_1\pr\lambda_1\pr}\ra&=&A(\om_1)\delta^3_{{\bf k}_1,{\bf k}_1\pr}\delta_{\lambda_1,\lambda_1\pr}, \ \ \ \om_1=|{\bf k}_1|c,\nonumber\\
\la \tilde{a}\rk \tilde{a}_{{\bf k}_1\pr\lambda_1\pr}\ra&=&0,
\label{eq109}
\eea
for the classical stochastic variables describing blackbody radiation, and likewise for the averages of the terms denoted by subscript 2. Here $A(\om)$ is a positive number defined by the first line of Eq.~\eqref{eq109}.

We have assumed as in the model of  Einstein and Hopf that the induced dipole is along the $z$ direction, and have therefore dealt with only the $z$ component of the electric field. To account for all three components of the electric field we replace the polarization sums by
\be
\sum_{\lambda_1}\big[e^2_{{\bf k}_1\lambda_1x}+e^2_{{\bf k}_1\lambda_1y}+e^2_{{\bf k}_1\lambda_1z}\big]\sum_{\lambda_2}\big[e^2_{{\bf k}_2\lambda_2x}+e^2_{{\bf k}_2\lambda_2y}+e^2_{{\bf k}_2\lambda_2z}\big]=\sum_{\lambda_1}{\bf e}^2_{{\bf k}_1\lambda_1}\sum_{\lambda_2}{\bf e}^2_{{\bf k}_2\lambda_2}=4,
\ee
since ${\bf e}_{{\bf k}\lambda}$ is a unit vector, and then average over the contributions from the three electric field components. Taking $V\rightarrow\infty$ in the usual fashion, we then have
\bea
\la\Delta p^2\ra&=\frac{1}{3}\times&\frac{32(2\pi\hbar)^2}{(8\pi^3)^2c^8}\frac{4\pi}{3}4\pi\int_0^{\infty}d\ok\ok^5\int_0^{\infty}d\okk\okk^3\frac{\sin^2\bkt{\frac{1}{2}(\ok-\okk)\Delta t}}{(\ok-\okk)^2}\nonumber\\
&&\mbox{}\times A(\ok)A(\okk)\abs{\alpha(\okk)}^2.
\label{eq110}
\eea
The factors of $4\pi/3$ and $4\pi$ result from the angular integration over ${\bf k}_1$  and ${\bf k}_2$, respectively in Eq.~\eqref{eq108}.
The term involving $\ok-\okk$ in Eq.\eqref{eq110} is sharply peaked at $\ok=\okk$ compared with the remaining factors in the integrand. We therefore evaluate these other factors at $\okk=\ok$ and use
\be
\int_0^{\infty}d\okk\frac{\sin^2\bkt{\frac{1}{2}(\okk-\ok)\Delta t}}{(\okk-\ok)^2} 
\cong\frac{1}{2}\int_{-\infty}^{\infty}dx\frac{\sin^2{x\Delta t}}{x^2}=\pi\Delta t/2
\ee
in the remaining part of the $\okk$ integral to obtain
\be
\frac{\la\Delta p^2\ra}{\Delta t}=\frac{16\hbar^2}{9\pi c^8}\int_0^{\infty}d\ok \ok^8A^2(\ok)|\alpha(\ok)|^2.
\ee
Using again the optical theorem Eq.~\eqref{eq206}, we have finally
\be
\frac{\la\Delta p^2\ra}{\Delta t}=\frac{8\hbar^2}{3\pi c^5}\int_0^{\infty}d\om\om^5A^2(\om)\alpha_I(\om)
\label{eq111}
\ee
for the momentum diffusion constant.

The expression Eq.~\eqref{eq111} was obtained from a classical stochastic treatment of the field. In the quantum treatment of the field $\tilde{a}\rk$ and $\tilde{a}^*\rk$ in 
Eq.~\eqref{eq108}  are replaced by photon annihilation and creation operators ${a}\rk$
and ${a}^{\dag}\rk$, respectively. Then the quantized-field expression for the momentum diffusion constant can be obtained by replacing classical ensemble averages by quantum expectation values:
\bea
\la \tilde{a}\rk \tilde{a}^*\rk\ra&\rightarrow& \la {a}\rk {a}^{\dag}\rk\ra=n(\ok)+1,\nonumber\\
\la \tilde{a}^*\rkk \tilde{a}\rkk\ra&\rightarrow& \la {a}^{\dag}\rkk {a}\rkk\ra=n(\okk),\nonumber\\ 
\la \tilde{a}^*\rk \tilde{a}\rk\ra&\rightarrow&\la {a}^{\dag}\rk {a}\rk\ra=n(\ok),\nonumber\\
\la \tilde{a}\rkk \tilde{a}^*\rkk\ra&\rightarrow&\la {a}\rkk {a}^{\dag}\rkk\ra=n(\okk)+1.
\label{eq112}
\eea
Then, following essentially the same arguments that led from Eq.~\eqref{eq108} to Eq.~\eqref{eq111}, we obtain the quantized-field expression for the momentum diffusion constant:
\be
\frac{\la \Delta p^2\ra}{\Delta t}=\frac{8\hbar^2}{3\pi c^5}\int_0^{\infty}d\om\om^5\alpha_I(\om)[n^2(\om)+n(\om)],
\label{eq113}
\ee
which differs from Eq.~\eqref{eq210} by the Bose--Einstein factor $n(\om)+1$. This is our general expression for the momentum diffusion constant of a small particle in blackbody radiation. We note that, with a bit more algebra, it follows directly of course from a fully quantized-field calculation; this involves the symmetrization of  $\Delta p$ in Eq.~\eqref{eq107} to make it manifestly Hermitian, and taking  $\tilde{a}\rk$ and $\tilde{a}^*\rkk$ to be  the operators ${a}\rk$ and ${a}^{\dag}\rkk$, respectively. 
The physical significance of the $n^2(\om)$ and $n(\om)$ contributions to the momentum diffusion constant is discussed further in Section \ref{sec:discussion}.

Defining a wave number $K=p/\hbar$ associated with a de Broglie wavelength $\lambda_{{\rm dB}}=2\pi/K$, and using again the optical theorem Eq.~\eqref{eq206}, we can express Eq.~\eqref{eq113} as
\bea
\frac{\la \Delta K^2\ra}{\Delta t}
&=&\frac{16}{9\pi c^8}\int_0^{\infty}d\om\om^8|\alpha(\om)|^2[n^2(\om+n(\om)]\nonumber\\
&=&\frac{8}{3\pi c^5}\int_0^{\infty}d\om\om^5\alpha_I(\om)[n^2(\om)+n(\om)],
\label{eqsc1}
\eea
As discussed below,  $(\la \Delta K^2\ra/\Delta t)\Delta x^2$ is closely related to $\Lambda\Delta x^2$, the measure of the rate at which interference, or coherent superposition, is lost between  two localized wave packets separated by a distance $\Delta x$. $\Lambda$ here is  the ``scattering constant"  appearing in the theory of collisional decoherence due to thermal photon scattering. 

\subsubsection{Electron in blackbody radiation}
 Consider as an example an electron in blackbody radiation. From the classical Abraham-Lorentz equation of motion 
\be
m_e\ddot{\br}-m_e\tau_e\stackrel{...}{\br}=\bE,
\ee
we deduce a ``polarizability"
\be
\alpha(\om)=-\frac{e^2/m_e}{\om^2+i\tau_e\om^3}.
\ee
Since $\tau_e=2e^2/3m_ec^3\cong 6.3\times 10^{-24}$ s, we approximate $\alpha_I(\om)$  by 
\be
\alpha_I(\om)=\frac{e^2\tau_e}{m_e\om}=\frac{c\sigma_T}{4\pi\om},
\label{eq12b}
\ee
where $\sigma_T=(8\pi/3)(e^2/m_ec^2)^2$ is the Thomson scattering cross section. Then, from Eq.~\eqref{eq113} with $n(\om)=(e^{\hbar\om/k_BT}-1)^{-1}$,
\be
\frac{\la\Delta p_e^2\ra}{\Delta t}=\frac{16\hbar^2}{9\pi c^4}\bkt{\frac{e^2}{m_ec^2}}^2\int_0^{\infty}d\om\frac{\om^4e^{\hbar\om/k_BT}}{(e^{\hbar\om/k_BT}-1)^2}=\frac{64\pi^3}{135}\bkt{\frac{e^2}{m_ec^2}}^2\frac{(k_BT)^5}{\hbar^3c^4},
\label{ox1}
\ee
exactly as obtained by Oxenius \cite{oxenius} in a different approach based directly on Thomson scattering. $\la\Delta K_e^2\ra/\Delta t$, aside from a numerical prefactor due to a different averaging procedure, has the same form as the decoherence rate for free electrons obtained by Joos and Zeh \cite{joos}.

The role of the Bose--Einstein factor in this example turns out to be very small. If we replace $n^2(\om)+n(\om)$ by $n(\om)$ in Eq.~\eqref{eqsc1}  we obtain \eqn{ \frac{\avg{\Delta p_e^2}}{\Delta t}\approx \frac{128}{3\pi}\zeta\bkt{5} \bkt{\frac{e^2}{m_e c^2}}^2\frac{\bkt{k_B T}^5}{\hbar^3 c^4},} with $\zeta(5)\approx1.0369$, which is about 96\% of the complete result with both terms retained. $n^2(\om)$ is much larger than $n(\om)$ at low frequencies, but this dominance is suppressed by the $\om^5$ factor in the integrand of Eq. 
(\ref{eqsc1}); this is due in part to the fact that the field mode density decreases rapidly with decreasing frequency. The suppression of the $n^2(\om)$ contribution is even more pronounced in the following example.
\subsubsection{Dielectric sphere in blackbody radiation}
For a small dielectric sphere of radius $a$, 
\be
\alpha(\om)=\bkt{\frac{\eps-1}{\eps+2}}a^3,
\label{eqsc2}
\ee
and, if $n(\om)=(e^{\hbar\om/k_BT}-1)^{-1}$ and the variation of the permittivity $\eps$ with $\om$ is negligible,
\eqn{
\frac{\la \Delta K_s^2\ra}{\Delta t}=&\frac{16}{9\pi c^8}a^6\abs{\frac{\eps-1}{\eps+2}}^2\int_0^{\infty}d\om\om^8\frac{e^{\hbar\om/k_BT}}{(e^{\hbar\om/k_BT}-1)^2}= \frac{1024\pi^7 }{135}a^6c \abs{\frac{\epsilon-1}{\epsilon+ 2}}^2 \bkt{\frac{k_B T }{\hbar c}}^9.
\label{eqsc3}
}
If as above we retain only the term $n(\om)$ in Eq.~\eqref{eqsc1}, we obtain, independently of the temperature, 99.8\% of the full result, i.e.,
\bea
\frac{\la\Delta K_s^2\ra}{\Delta t}&\cong&\frac{16}{9\pi c^8}a^6\abs{\frac{\eps-1}{\eps+2}}^2\int_0^{\infty}d\om\om^8\frac{1}{e^{\hbar\om/k_BT}-1}\nonumber\\
&=&8!\frac{16}{9\pi}\zeta(9)a^6c\abs{\frac{\eps-1}{\eps+2}}^2\bkt{\frac{k_BT}{\hbar c}}^9\equiv 2\Lambda,
\label{eqsc4}
\eea
$\zeta(9)\cong 1.002$. For this example $\Lambda$ as defined by Eq.~\eqref{eqsc4} is the scattering constant derived, for example, in the book by Schlosshauer \cite{joos, schloss}.

The momentum diffusion rate we calculated is a mean-square ensemble average over momentum fluctuations and makes no reference to entanglement with the environment that is generally understood to be the hallmark of decoherence \cite{schloss}. $\Lambda$ as defined in our Eq. (\ref{eqsc4}) is nevertheless consistent with rigorous calculations of the decoherence rate \cite{diosi, horn,adler,schloss}. In a similar vein Adler \cite{adler} has compared a diffusion rate with results of ``decoherence-based" calculations of collisional decoherence to provide more generally an ``independent cross-check" of those calculations. Such a correspondence between the center-of-mass decoherence rate and the momentum diffusion has also been deduced from the variance of momentum as governed by the position-localization-decoherence master equation \cite{Oriol11}.

\section{\label{sec:decoh} Field Interference and Decoherence}
Consider two point dipoles $\bd_1$  and $\bd_2$ at fixed positions $\bx_1$ and $\bx_2$. The interaction Hamiltonian  may be taken for our purposes to be 
\be
H_I=-\bd_1\cdot\bE(\bx_1,t)-\bd_2\cdot\bE(\bx_2,t).
\ee
The electric field operator at a point $\bx$ may be expanded in plane-wave mode functions as in Eq.~\eqref{eq104}, but here $a\rmk(t)$ and $a^{\dag}\rmk(t)$ are photon annihilation and creation operators:
\be
\bE(\br,t)=i\sum\rmk\bkt{\frac{2\pi\hbar\om }{V}}^{1/2}\big[a\rmk(t)e^{i\bk\cdot\bx}-a^{\dag}\rmk(t)e^{-i\bk\cdot\bx}\big]\bek .
\ee
From the Heisenberg equation of motion
\be
i\hbar\dot{a}\rmk=[a\rmk,H_F]+[a\rmk,H_I]
\ee
with $H_F=\sum\rmk a^{\dag}\rmk a\rmk$ and the commutation relations for the photon operators, it follows that 
\be
a_{\bk\lambda i}(t)=a\rmk(0)e^{-i\om t}-\bkt{\frac{2\pi\om }{\hbar V}}^{1/2}\sum_i e^{-i\bk\cdot\bx_i}\int_0^tdt\pr \big[\bd_i(t\pr)\cdot\bek\big]e^{i\om (t\pr-t)}.
\label{eq100}
\ee

\begin{figure}
    \centering
    \includegraphics[width = 2.25 in]{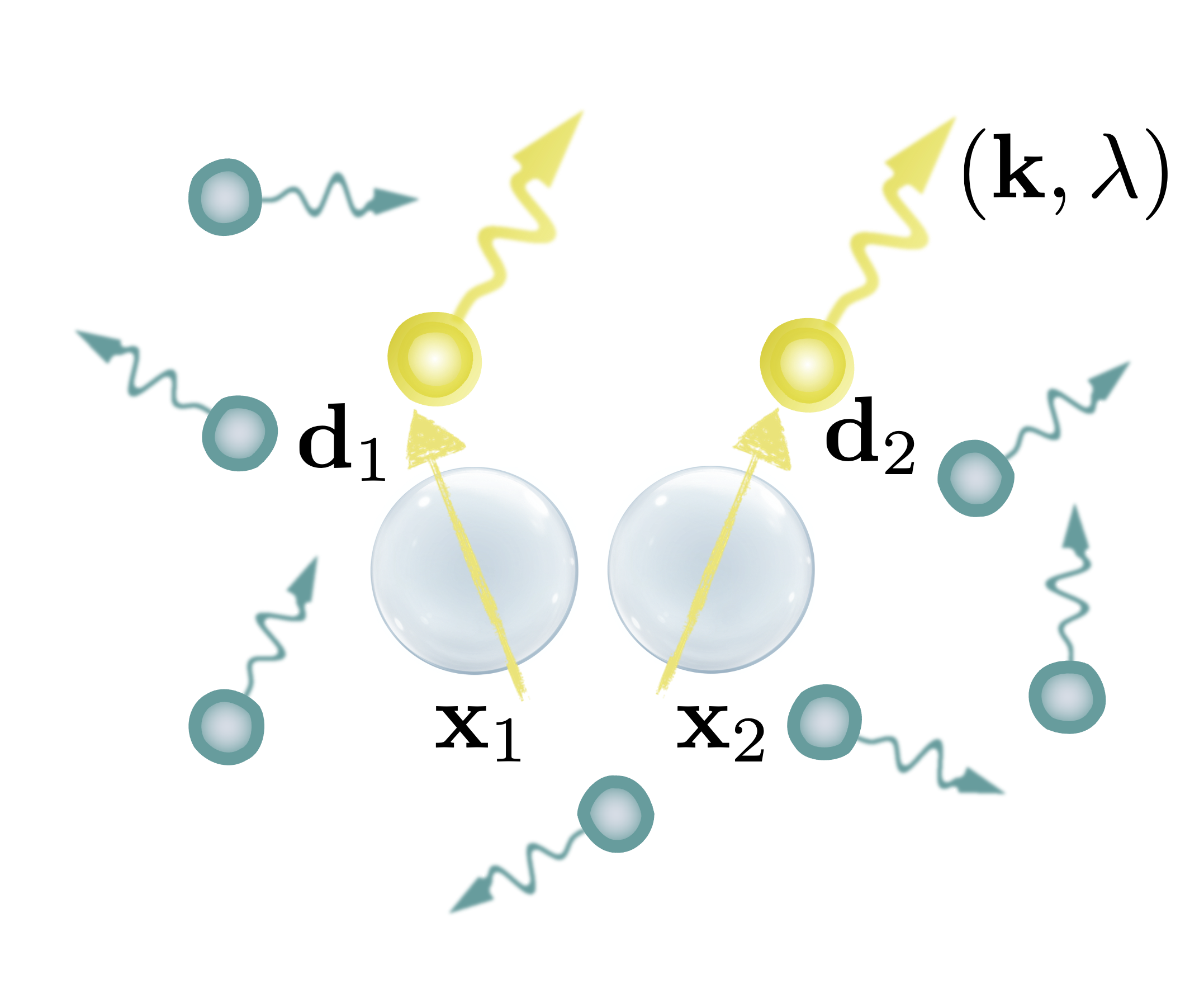}
    \caption{Schematic representation of two dipoles located at positions $\mathbf{x}_{1}$ and $\mathbf{x}_{2}$, interacting with a thermal field. We consider the rate $R$ of the average number of photons emitted from the  dipole moments $\mathbf{d}_{1}$ and $\mathbf{d}_{2}$ induced by the blackbody field. }
    \label{Sch2}
\end{figure}
We define a rate of change of the average number of photons radiated by the dipole moments induced at $\bx_1$ and $\bx_2$:
\be
R=\frac{d}{dt}\sum\rmk\big\la \big[a^{\dag}\rmk(t)-a^{\dag}\rmk(0)e^{i\om t}\big]\big[a\rmk(t)-a\rmk(0)e^{-i\om t}\big]\big\ra.
\ee
From Eq.~\eqref{eq100}, 
\be
R=\frac{2\pi}{\hbar V}\times 2{\rm Re}\sum_{i,j}\sum_{\bk}\om e^{i\bk\cdot(\bx_i-\bx_j)}\int_0^tdt\pr e^{i\om (t\pr-t)}\sum_{\lambda}\avg{\sbkt{\bd_i^{\dag}(t)\cdot\bek}\sbkt{\bd_j(t\pr)\cdot\bek}} .
\label{eqpp}
\ee

Consider first
\be
X_{ij}\equiv\int_0^tdt\pr e^{i\om (t\pr-t)}\sum_{\lambda}\Big\la[\bd_i^{\dag}(t)\cdot\bek][\bd_j(t\pr)\cdot\bek]\Big\ra ,
\label{eq}
\ee
where $\bd_i(t)$ is the dipole moment induced by the blackbody field at $\bx_i$:
\bea
\bd_i(t)\cdot\bek&=&i\sum_{\bk\pr\lambda\pr}\bkt{\frac{2\pi\hbar\om\pr}{V}}^{1/2}\Big[\alpha(\om{\pr})a_{\bk\pr\lambda\pr}(0)e^{-i\om{\pr}t}e^{i\bk\pr\cdot\bx_i}
-\alpha^*\bkt{\om{\pr}}a^{\dag}_{\bk\pr\lambda\pr}(0)e^{i\om{\pr}t}e^{-i\bk\pr\cdot\bx_i}\Big]\nonumber\\
&&\mbox{}\times{\bf e}_{\bk\pr\lambda\pr}\cdot\bek.
\eea
For $X_{ij}$ we obtain
\bea
X_{ij}&=&\sum_{\bk\pr\lambda\pr}\Big(\frac{2\pi\hbar\om{\pr}}{V}\Big)\Big\{|\alpha(\om{\pr})|^2
\Big[\la a_{\bk\pr\lambda\pr}^{\dag}(0)a_{\bk\pr\lambda\pr}(0)\ra e^{-i\bk\pr\cdot(\bx_i-\bx_j)}\int_0^tdt^{\prime}e^{i(\om -\om{\pr})(t\pr-t)}\nonumber\\
&&\mbox{}+\la a_{\bk\pr\lambda\pr}(0)a_{\bk\pr\lambda\pr}^{\dag}(0)\ra e^{i\bk\pr\cdot(\bx_i-\bx_j)} \int_0^tdt^{\prime}e^{i(\om +\om{\pr})(t\pr-t)}\Big]
|\hat{e}\rmk\cdot\hat{e}_{\bk\pr\lambda\pr}|^2\Big\}.
\label{xequation}
\eea
We have used the expectation values
\be
\big\la a^{\dag}_{\bk\pr\lambda\pr}(0)a_{\bk\dpr\lambda\dpr}(0)\big\ra=n_{\bk\pr\lambda\pr}\delta^3_{\bk\pr\bk\dpr}\delta_{\lambda\pr\lambda\dpr}=n_{k\pr}\delta^3_{\bk\pr\bk\dpr}\delta_{\lambda\pr\lambda\dpr}, \ \ \ \big\la a_{\bk\pr\lambda\pr}(0)a_{\bk\dpr\lambda\dpr}(0)\big\ra=0
\ee
for the blackbody field. From the form of the time integrals in Eq. (\ref{xequation}) it is seen that only the normally ordered expectation value will contribute to the final result, and therefore we write
\be
X_{ij}=\frac{2\pi\hbar}{V}\sum_{\bk\pr}\om{\pr}|\alpha(\om{\pr})|^2n_{k\pr} e^{-i\bk\pr\cdot(\bx_i-\bx_j)}\sum_{\lambda}\sum_{\lambda\pr}|{\bf e}_{\bk\pr\lambda\pr}\cdot\bek|^2\int_0^tdt\pr e^{i(\om -\om{\pr})(t\pr-t)},
\ee
or, since
\be
\sum_{\lambda}\sum_{\lambda\pr}|{\bf e}_{\bk\pr\lambda\pr}\cdot\bek|^2=1+(\bk\cdot\bk\pr)^2/(kk\pr)^2,
\ee
\be
X_{ij}=\frac{2\pi\hbar}{V}\frac{V}{8\pi^3}\int d^3k\pr \om{\pr}|\alpha(\om{\pr})|^2 n_{k\pr}e^{-i\bk\pr\cdot(\bx_i-\bx_j)}\sbkt{1+(\bk\cdot\bk\pr)^2/(kk\pr)^2}\int_0^tdt\pr e^{i(\om -\om{\pr})(t\pr-t)}
\ee
in the mode continuum limit. Therefore, 
\bea
R&=&\frac{4\pi}{\hbar V}\frac{2\pi\hbar}{V}\frac{V}{8\pi^3}\frac{V}{8\pi^3}\sum_{i,j}\int d^3k\om \int d^3k\pr\om{\pr}|\alpha(\om{\pr})|^2n_ke^{i(\bk-\bk\pr)\cdot(\bx_i-\bx_j)}\sbkt{1+(\bk\cdot\bk\pr)^2/(kk\pr)^2}\nonumber\\&&\mbox{}\times\int_0^tdt\pr\cos\sbkt{(\om -\om{\pr})(t\pr-t)}.
\eea
For $t\rightarrow\infty$,
\be
\int_0^tdt\pr\cos\sbkt{(\om -\om{\pr})(t\pr-t)}\rightarrow \pi\delta(\om -\om{\pr})=\frac{\pi}{c}\delta(k-k\pr),
\ee
and 
\bea
R&=&\sum_{i,j}\frac{c}{8\pi^3}\int d^3kk\int d^3k\pr k\pr|\alpha(\om{\pr})|^2n_ke^{i(\bk-\bk\pr)\cdot(\bx_i-\bx_j)}\big[1+(\bk\cdot\bk\pr)^2/(kk\pr)^2\Big]\delta(k-k\pr)\nonumber\\
&=&\sum_{i,j}\frac{c}{8\pi^3}\int dkk^6|\alpha(\om)|^2n_k\int d\Om_{\hat{\bk}}\int d\Om_{\hat{\bk}\pr}e^{ik(\hat{\bk}-\hat{\bk}\pr)\cdot(\bx_i-\bx_j)}\big(1+\cos^2\theta\big)\nonumber\\
&\equiv&\sum_{i,j} R_{ij},
\eea
where $\hat{\bk}$ is a unit vector in the direction of $\bk$, $\theta$ is the angle between $\hat{\bk}$ and $\hat{\bk}\pr$, and $d\Om_{\hat{\bk}}$ is a differential element of solid angle about $\hat{\bk}$ 
($\int d^3k=\int dkk^2\int d\Om_{\hat{\bk}}$).

We rewrite
\be
R_{12}=\frac{c}{8\pi^3}\int dkk^6|\alpha(\om)|^2n_k\int d\Om_{\hk}\int d\Om_{\hk\pr}e^{ik(\hk-\hk\pr)\cdot(\bx_i-\bx_j)}\big(1+\cos^2\theta\big)
\ee
using $n_k=(e^{\hbar kc/k_BT}-1)^{-1}$ and the differential cross section for Rayleigh scattering:
\be
|f(\bk,\bk^{\prime})|^2=\frac{d\sigma}{d\Om}=k^4|\alpha(\om)|^2\frac{1}{2}(1+\cos^2\theta).
\ee
Thus
\be
R_{12}=\frac{c}{4\pi^3}\int dk\frac{k^2}{e^{\hbar kc/k_BT}-1}\int d\Om_{\hk}\int d\Om_{\hk\pr}e^{ik(\hk-\hk\pr)\cdot(\bx_i-\bx_j)}\big|f(k\hat{\bk},k\hat{\bk\pr})\big|^2.
\ee
We express this in terms of $q=\hbar k$, $v(q)=c$, and 
\be
dq\rho(q)=dk\rho(k)=\frac{k^2dk/\pi^2}{e^{\hbar kc/k_BT}-1},
\ee
and, to compare with Reference \cite{schloss}, for instance, we also change notation, writing $\bx_1-\bx_2=\bx-\bx\pr$, $d\Om_{\hk}d\Om_{\hk\pr}=d\hn d\hn\pr$, and replacing $\big|f(k\hk,k\hk\pr)\big|^2$ by
$\big|f(q\hn,q\hn\pr)\big|^2$. 
Then
\be
R_{12}=\int dq\rho(q)v(q)\int\frac{d\hn d\hn\pr}{4\pi}e^{iq(\hn-\hn\pr)\cdot(\bx-\bx\pr)/\hbar}\big|f(q\hn,q\hn\pr)\big|^2,
\ee
and similarly
\be
R_{11}=\int dq\rho(q)v(q)\int\frac{d\hn d\hn\pr}{4\pi}\big|f(q\hn,q\hn\pr)\big|^2.
\ee

The photon scattering rate $R_{12}$, as opposed to $R_{11}$  involves interference between the fields of the two dipoles. The difference $R_{11}-R_{12}$ is therefore a measure of the rate at which this interference decreases with increasing separation of the dipoles. In other words, it characterizes the loss of spatial coherence between the two dipole fields. In fact 
\be
F(\bx-\bx\pr) \equiv R_{11}-R_{12}=\int dq\rho(q)v(q)\int\frac{d\hn d\hn\pr}{4\pi}\Big(1-e^{iq(\hn-\hn\pr)\cdot(\bx-\bx\pr)/\hbar}\Big)\big|f(q\hn,q\hn\pr)\big|^2
\label{interfere}
\ee
happens to have exactly the form of the more general collisional decoherence rate \cite{diosi, horn,adler,schloss}. In the long-wavelength limit the decay rate of the reduced density matrix in the case of a dielectric sphere reduces to $-\Lambda(\bx-\bx\pr)^2$, where $\Lambda$  is defined in Eq.~\eqref {eqsc4}. The ``decoherence factor''  $F\bkt{\bx - \bx'}$, as it appears in the theory of collisional decoherence  \cite{diosi, horn,adler,schloss}, represents the overlap  between the states of environmental scatterers that scatter off the positions $\bx$ and $\bx'$ of the particle's center of mass, denoting their distinguishability.
 Though, it is to be noted that the scattering model of decoherence considers two points $\bx $ and $ \bx'$ located within the same particle whose center of mass is spatially delocalized; in the above calculation $\bx$ and $\bx'$ correspond to the location of two distinct dipoles, that are not necessarily constrained to constitute the same particle. 

\section {\label{sec:drag} Drag Force}
As noted in the Introduction, Einstein and Hopf \cite{eh1,eh2} and later Einstein \cite{ein1} showed that there is a drag force $F=-\xi m v$ on a polarizable particle moving with velocity $v$ ($v\ll c$) with respect to a frame in which there is an isotropic (blackbody) field of spectral energy density $\rho(\om)$. A more general expression for $\xi$ was derived much later \cite{mkr}, and more recently the drag force has been derived for relativistic particles \cite{dedkov, pieplow,volo,kim}. We now derive the relativistic form of the drag force based in part on ``energy kicks" $\hbar\om$, in the same heuristic spirit as in the treatment of momentum kicks in Section \ref{sec:kicks}. The particle is assumed to have a uniform velocity $v$ along the $x$ axis in the laboratory frame S, and we are interested in the force $F_x$ on the particle in this frame. The Lorentz transformations relating the spacetime coordinates and the momenta 
$(p_x,p_x\pr$) and energies ($\mc{E},\mc{E}\pr$) in the laboratory frame and the particle's rest frame S$'$ are
\bea
x\pr&=&\gamma(x-vt),\\
t\pr&=&\gamma\bkt{t-\frac{vx}{c^2}},\\
p_x\pr&=&\gamma\bkt{p_x-\frac{v}{c^2}\mc{E}},\\
\mc{E}\pr&=&\gamma\bkt{\mc{E}-vp_x},
\label{lor1}
\eea
where $dx/dt=v$ and as usual $\gamma=(1-v^2/c^2)^{-1/2}$. These transformations relate the force $F_x$ in S to the force $F_x\pr$ in S$'$:
\be
F_x=\frac{dp_x}{dt}=\frac{d}{dt\pr}\sbkt{\gamma\bkt{p_x\pr+\frac{v}{c^2}\mc E\pr}}\frac{dt\pr}{dt}=\frac{dp_x\pr}{dt\pr}+
\frac{v}{c^2}\frac{d\mc E\pr}{dt\pr}
=F_x\pr+\frac{v}{c^2}\frac{d\mc E\pr}{dt\pr}.
\label{lor2}
\ee

The force on the particle has two contributions, $F_{\rm ind}$ resulting from the dipole moment induced by the fluctuating blackbody field, and $F_{\rm dd}$ resulting from the dipole's own radiation field and fluctuations \cite{dedkov, pieplow,volo,kim}. $F_{\rm ind}$ is the force originally considered by Einstein and Hopf \cite{ein1,eh1,eh2}, while $F_{\rm dd}$ is needed to obtain the fully relativistic expression for the force that has been of interest in recent work \cite{dedkov, pieplow,volo,kim}. 

\subsection{Calculation of the force $F_{\rm ind}$}
In Section \ref{sec:induced} we calculated the force when the electric field is along a single direction $z$ and then exploited the isotropy of the field to include the contributions from all three components of the field. The situation is more complicated for a particle in motion because in its reference frame (S$'$) the fields are different in different directions and are related to those in S by
\bea
E_x\pr(\bx\pr,t\pr)&=&E_x(\bx,t),\nonumber\\
E_y\pr(\bx\pr,t\pr)&=&\gamma\sbkt{E_y(\bx,t)-\frac{v}{c}B_z(\bx,t)},\nonumber\\
E_z\pr(\bx\pr,t\pr)&=&\gamma\sbkt{E_z(\bx,t)+\frac{v}{c}B_y(\bx,t)},
\label{fieldtrans}
\eea
along with the corresponding transformations for the magnetic fields. The force $d_0\pa E_z/\pa x$ given by (\ref{eq1033}) applies when the dipole 
moment $d_0$ ($d_0=d_z$) and the electric field both have only $z$ components. More generally the force along the $x$ direction is
\be
F_x=d_x\frac{\pa E_x}{\pa x}+d_y\frac{\pa E_y}{\pa x}+d_z\frac{\pa E_z}{\pa x},
\label{ff1}
\ee
as is easily shown. For the quantum calculation it will be convenient to consider first a single field mode polarized along the $x$ direction:
\bea
E\pr(\bx\pr,t\pr)&=&E_x(\bx,t)=i\bkt{\frac{2\pi\hbar\om }{V}}^{1/2}\sbkt{a\rmk e^{-i(\om t-\bk\cdot\bx)}-{a}^{\dag}\rmk e^{i(\om t-\bk\cdot\bx)}}e_{\bk\lambda x}\nonumber\\
&=&i\bkt{\frac{2\pi\hbar\om }{V}}^{1/2}\sbkt{a\rmk e^{-i(\om_{\bk\pr} t\pr-\bk\pr\cdot\bx\pr)}-{a}^{\dag}\rmk 
e^{i(\om_{\bk\pr} t\pr-\bk\pr\cdot\bx\pr)}}e_{\bk\lambda x}.
\label{ff2}
\eea
We have used the fact that $e^{i(\om t-\bk\cdot\bx)}=e^{i(\om_{\bk}\pr t\pr-\bk\pr\cdot\bx\pr)}$ since we will require $\pa E_x\pr/\pa x\pr$. The primed and unprimed frequencies and wave vectors are related by
\bea
\om{\pr}&=&\gamma(\om -vk_x),\nonumber\\
k_x\pr&=&\gamma\bkt{k_x-\frac{v}{c^2}\om }, \ \ \ k_y\pr=k_y, \ \ \ k_z\pr=k_z.
\label{ff3}
\eea
The induced dipole moment $d_x\pr(\bx\pr,t\pr)$ is given by
\be
d_x\pr(\bx\pr,t\pr)=i\bkt{\frac{2\pi\hbar\om }{V}}^{1/2}\sbkt{\alpha(\om{\pr})a\rmk e^{-i(\om{\pr} t\pr-\bk\pr\cdot\bx\pr)}-\alpha^*(\om{\pr}){a}^{\dag}\rmk 
e^{i(\om{\pr} t\pr-\bk\pr\cdot\bx\pr)}}e_{\bk\lambda x},
\label{ff4}
\ee
since $\alpha(-\om_{\bk\pr})=\alpha^*(\om_{\bk\pr})$.

Consider first the force
\be
[F\pr_{\rm ind}]_{xx}=\frac{1}{2}\avg{ d'_x\frac{\pa E'_x}{\pa x'}+\frac{\pa E'_x}{\pa x'}d'_x}={\rm Re}\avg{ d'_x\frac{\pa E'_x}{\pa x'}},
\label{ff5}
\ee
corresponding quantum mechanically to the first term in (\ref{ff1}). The expectation value is over the thermal equilibrium state of the field. It follows from 
(\ref{ff2}) and {\ref{ff4}) that
\be
[F\pr_{\rm ind}]_{xx}=\frac{4\pi\hbar\om }{V}k_x\pr\alpha_I(\om_{\bk}\pr)n(\om )e^2_{\bk\lambda x},
\label{ff6}
\ee
where we have defined $n(\om )=\la{a}^{\dag}\rmk a\rmk\ra$ for the (unpolarized) blackbody field and used the fact that $\la a_{\bk_1\lambda_1}{a}_{\bk_2\lambda_2}\ra=0$ for this field. Next we add up the contributions to $F_{1x}\pr$ from all field modes: 
\bea
[F\pr_{\rm ind}]_{xx}&\rightarrow&\frac{V}{8\pi^3}\int d^3k\bkt{\frac{4\pi\hbar\om }{V}}k_x\pr\alpha_I(\om_{\bk\pr})n(\om )\sum_{\lambda=1}^2e^2_{\bk\lambda x}\nonumber\\
&=&\frac{\hbar}{2\pi^2}\int d^3k\om k_x\pr\bkt{1-\frac{k_x^2}{k^2}}\alpha_I(\om_{\bk\pr})n(\om ).
\label{ff7}
\eea

To compare with previous work we wish to write this force as an integral over the primed variables. To this end we note from the Jacobian that follows from Eq. (\ref{ff3}) that
\be
d^3k=\gamma\bkt{1+\frac{v}{c}\frac{k_x\pr}{k\pr}}d^3k\pr,
\ee
and that, again from Eq. (\ref{ff3}),
\bea
1-\frac{k_x^2}{k^2}&=&\frac{1}{\gamma^2}\frac{1-{k_x\pr}^2/{k\pr}^2}{\big(1+\frac{v}{c}\frac{k_x\pr}{k\pr}\big)^2},\nonumber\\
\om &=&\gamma(\om{\pr}+vk_x\pr).
\eea
It then follows that
\bea
[F\pr_{\rm ind}]_{xx}&=&\frac{\hbar c}{2\pi^2}\int d^3k\pr k\pr k_x\pr\bkt{1-\frac{{k_x\pr}^2}{{k\pr}^2}}\alpha_I(\om{\pr})n(\gamma(\om{\pr}+vk_x\pr))\nonumber\\
&=&\frac{\hbar c}{2\pi^2}\int d^3k kk_x\bkt{1-\frac{k^2_x}{k^2}}\alpha_I(\om )n(\gamma(\om +vk_x)),
\eea
where in the last step we simply replace primed dummy integration variables by unprimed ones.

The remaining contributions to $F_{1x}\pr$ corresponding to the last two terms in Eq. (\ref{ff1}), are found similarly. We simply state the results of the straightforward calculations:
\be
[F\pr_{\rm ind}]_{xy}=\frac{\hbar c}{2\pi^2}\int d^3k kk_x\bkt{1-\frac{k^2_y}{k^2}}\alpha_I(\om )n(\gamma(\om +vk_x)),
\ee
\be
[F\pr_{\rm ind}]_{xz}=\frac{\hbar c}{2\pi^2}\int d^3k kk_x\bkt{1-\frac{k^2_z}{k^2}}\alpha_I(\om )n(\gamma(\om +vk_x)).
\ee
Summing up all the components, one obtains:
\bea
F\pr_{\rm ind}&=&[F\pr_{\rm ind}]_{xx}+[F\pr_{\rm ind}]_{xy}+[F\pr_{\rm ind}]_{xz}\nonumber\\
&=&\frac{\hbar c}{2\pi^2}\int d^3k kk_x\bkt{1-\frac{k_x^2}{k^2}+1-
\frac{k_y^2}{k^2}+1-\frac{k_z^2}{k^2}}\alpha_I(\om)n(\gamma(\om+vk_x))\nonumber\\
&=&\frac{\hbar c}{\pi^2}\int d^3k kk_x\alpha_I(\om)n(\gamma(\om+vk_x))
\label{ff8}
\eea

$F\pr_{\rm ind}$ is by definition the average force in the particle frame due to the fluctuating dipole moment induced by the blackbody field in the particle frame. There is no contribution to this force from the dipole's fluctuations from radiation reaction. Furthermore there is no contribution to $F\pr_{\rm ind}$ of the type $(v/c^2)d\mc E\pr/dt\pr=(v/c^2)(\bF\cdot{\bf u})$, since the velocity ${\bf u}=0$ in the particle frame. Equation (\ref{lor2}) therefore implies that $F_{\rm ind}$, the force in the laboratory frame due to the fluctuation dipole moment induced by the blackbody field, is identical to the force $F\pr_{\rm ind}$ in the particle frame:
\be
F_{\rm ind}=\frac{\hbar c}{\pi^2}\int d^3k kk_x\alpha_I(\om)n(\gamma(\om+vk_x)),
\label{forceinduced}
\ee
where $n(\om)=(e^{\hbar\om/k_bT}-1)^{-1}$ and $T$ is the temperature in the laboratory frame.

Before proceeding we consider the nonrelativistic limit of Eq. (\ref{ff8}), using
\be
n(\gamma(\om+vk_x))\cong n(\om)+vk_x\frac{dn}{d\om}=n(\om)-vk_x\frac{3\pi^2c^3}{\hbar\om^4}\sbkt{\rho(\om)-\frac{\om}{3}\frac{d\rho}{d\om}},
\ee
which follows from Eq. (\ref{pla}). Then, carrying out the integration in Eq. (\ref{ff8}) under this approximation, we obtain
\be
F_{\rm ind}=\frac{4\hbar}{3\pi c^5}v\int_0^{\infty}d\om\om^5\alpha_I(\om)\frac{dn}{d\om}
=-\frac{4\pi}{c^2}v\int_0^{\infty}d\om\om\alpha_I(\om)\sbkt{\rho(\om)-\frac{\om}{3}\frac{d\rho}{d\om}}\equiv -\xi m v
\label{nonreldrag}
\ee
for $v\ll c$. In this limit this is the total force on the moving particle in blackbody radiation in the laboratory frame. It has the basic form obtained by Einstein and Hopf \cite{eh1} and is equivalent to the more general expression obtained by Mkrtchian et al. \cite{mkr}. The obtained drag force can be physically understood as follows: the moving particle creates an asymmetric density of modes due to a relativistic aberration effect~\cite{QObook}, the preferential emission into forward modes results in a friction force due to recoil in the opposite direction.

One can consider the thermal drag for the previous examples of an electron and a dielectric sphere. The non-relativistic thermal drag coefficient for an electron in blackbody radiation can be obtained from Eq.~\eqref{nonreldrag} and Eq.~\eqref{eq12b} as 
\eqn{
\xi_e = \frac{32\pi^3\hbar  }{135 m_e} \bkt{\frac{e^2}{m_e c^2}}^2\bkt{\frac{k_B T}{\hbar c}}^4.
}

Similarly, the non-relativistic thermal drag on a dielectric sphere is found from Eq.~\eqref{nonreldrag} and Eq.~\eqref{eqsc2} to be
\eqn{
\xi_s = \frac{512 \pi^7 \hbar }{135 m_s} \abs{\frac{\eps-1}{\eps +2}}^2 \bkt{\frac{k_B T}{ \hbar c}}^8,
}
where $m_s $ represents the mass of the dielectric sphere.

We note that the friction coefficients ($\xi_{e}$ and $\xi_s$)  and the  momentum diffusion constants (Eq.~\eqref{ox1} and Eq.~\eqref{eqsc3}) for an electron and a dielectric sphere, respectively,  obey the fluctuation-dissipation relation
\be
\frac{\la\Delta p^2\ra}{\Delta t}=2m\xi k_BT
\label{fd}.
\ee
We will discuss this relation further in Section~\ref{sec:discussion}.

\subsection{Calculation of the force $F_{\rm dd}$}
$F_{\rm dd}$ is by definition the force due to the dipole's radiation reaction, and is therefore attributable to radiation by the dipole. It is given according to Eq. (\ref{lor2}) by 
$(v/c^2)d\mc E\pr/dt\pr$, with $d\mc E\pr/dt\pr$ the rate of change due to radiation of the particle's energy in its reference frame. $d\mc E\pr/dt\pr$ cannot of course depend on the velocity $v$, and so it must be just minus the energy absorption rate $R_{\rm abs}$ of the particle if it were in equilibrium with an isotropic Planck spectrum $\rho(\om)$ at the temperature in S$'$. For $R_{\rm abs}$ we write, in close analogy to our treatment of momentum kicks in 
Section \ref{sec:kicks},
\be
R_{\rm abs}=-d\mc E\pr/dt\pr=\int_0^{\infty}\hbar\om\big[u(\om)d\om\big]=
\int_0^{\infty}\hbar\om\big[c\rho(\om)/\hbar\om\big]\sigma(\om)d\om,
\label{lor3}
\ee
and so, from Eqs.~\eqref{eq204}, \eqref{pla}, and \eqref{eq206}, 
\be
F_{\rm dd}=-\frac{4v\hbar}{\pi c^5}\int_0^{\infty}d\om\om^4\alpha_I(\om)n(\om)=-\frac{4v\hbar}{\pi c^5}\int_0^{\infty}d\om\om^4\alpha_I(\om)
\frac{1}{e^{\hbar\om/k_BT\pr}-1},
\label{f2}
\ee
where $T\pr$ is the thermal equilibrium temperature in S$'$.

We note that a force of the form $(v/c^2)d\mc E\pr/dt\pr$ is consistent with the insightful analysis by Sonnleitner et al. \cite{steve} of a ``vacuum friction force" for an excited atom moving with uniform velocity $v$ in a vacuum. For example, for a two-level atom in its excited state at $t=0$, 
\be
F_x=(v/c^2)d\mc E\pr/dt\pr\cong (v/c^2)\frac{d}{dt}\sbkt{\hbar\om_0e^{-\Gamma t}}=-\frac{v}{c^2}\hbar\om_0\Gamma e^{-\Gamma t}
\ee
to order $v/c$, where $\om_0$ and $\Gamma$ are respectively the atom's transition frequency and spontaneous emission rate in S$'$ \cite{note3}.

The total force $F_x=F_{\rm ind}+F_{\rm dd}$ in the laboratory frame follows from Eqs. (\ref{lor2}), (\ref{ff8}), and (\ref{f2}). It takes the form given by Milton et al. \cite{kim}:
\bea
F_x&=&-\frac{2\hbar}{\pi\gamma^2v^2}\int_0^{\infty}d\om\om^4\alpha_I(\om)\int_{u_-}^{u_+}dy\bkt{y-\frac{1}{\gamma}}
\sbkt{\frac{1}{e^{\hbar\om/k_BT\pr}-1}-\frac{1}{e^{\hbar\om y/k_BT}-1}}\nonumber\\
&=&-\frac{\hbar}{\pi\gamma^2v^2}\int_0^{\infty}d\om\om^4\alpha_I(\om)\int_{u_-}^{u_+}dy\bkt{y-\frac{1}{\gamma}}
\big[\coth{(\hbar\om/2k_BT\pr)}-\coth{(\hbar\om y/2k_BT)}\big],\nonumber\\
\eea
where $u_+=\sqrt{\frac{1+v/c}{1-v/c}}$, $u_-=\sqrt{\frac{1-v/c}{1+v/c}}$ \cite{kim}.

\section{\label{sec:discussion} Discussion}

The average photon number $n(\om)$ in thermal equilibrium appears differently in the momentum diffusion constant, the decoherence rate, and the drag force. This is because the momentum diffusion constant has a fourth-order dependence on the electric field, whereas the decoherence rate and drag force have a second-order dependence.

The momentum diffusion constant we have obtained for a small polarizable particle in blackbody radiation, for example, involves the factor $n^2(\om)+n(\om)$. The two contributions correspond respectively to the ``wave" and ``particle" contributions to the Einstein fluctuation formula for blackbody radiation \cite{QObook}. But in the two examples we considered---a free electron and a dielectric sphere---the $n^2(\om)$ contribution is negligible and the momentum diffusion constant has effectively only a ``particle" contribution from the field. It then has the same form as the collisional decoherence rate $\Lambda$. 

In the case of decoherence due to collisions with gas molecules \cite{diosi,adler}, in contrast, no ``wave" or Bose--Einstein factor appears because the de Broglie wavelengths of the gas molecules are so small that quantum statistics are irrelevant.

Of course the $n^2(\om)$ and $n(\om)$ terms in the momentum variance and the drag force are both essential in determining the Planck spectrum. In thermal equilibrium the rate of change of kinetic energy due to momentum diffusion, $(1/2m)\la\Delta p^2\ra/\Delta t$, plus the rate of change $F_xv=-m\xi v^2=-2\xi\la \frac{1}{2}mv^2\ra=-\xi k_BT$ due to the drag force, must add to zero. ($m$ is the particle mass and $\xi$ is defined by Eq. (\ref{nonreldrag}).) Since for our purposes only the nonrelativistic momentum diffusion rate was needed, we consider here the nonrelativistic drag force. The condition for thermal equilibrium of the field, assuming thermal equipartition of the particles' kinetic energy, is then
\be
\frac{1}{2m}\frac{8\hbar^2}{3\pi c^5}\int_0^{\infty}d\om\om^5\alpha_I(\om)[n^2(\om)+n(\om)]+k_BT\frac{4\hbar}{3\pi c^5}\int_0^{\infty}d\om\om^5\alpha_I(\om)\frac{dn}{d\om}=0,
\label{fd1}
\ee
or
\be
\frac{dn}{d\om}=-\frac{\hbar}{k_BT}(n^2+n).
\label{ode}
\ee
This follows independently of $\alpha_I(\om)$, assuming $\alpha_I(\om)>0$ as implied by our use of the optical theorem for Rayleigh scattering. If we assume the form of the momentum diffusion constant obtained in Section \ref{sec:kicks}, we obtain
\be
\frac{dn}{d\om}=-\frac{\hbar}{k_BT}n
\ee
instead of Eq. (\ref{ode}), with the solution 
\be
n(\om)=(\rm const.) \times e^{-\hbar\om/k_BT}.
\ee
Together with Eq. (\ref{pla}), this implies the Wien form of the spectrum. If instead we assume the momentum diffusion constant $\propto n^2$ obtained with classical stochastic wave theory by Einstein and Hopf, we obtain
\be
\frac{dn}{d\om}=-\frac{\hbar}{k_BT}n^2
\ee
instead of Eq. (\ref{ode}), with the solution
\be
n(\om)=k_BT/\hbar\om,
\ee
when a constant of integration is chosen to give the Rayleigh--Jean spectrum. The solution of Eq. (\ref{ode}) is of course
\be
n(\om)=(e^{\hbar\om/k_BT}-1)^{-1},
\ee
when a constant interpretable in terms of a chemical potential is chosen so that the chemical potential is zero for photons, resulting in the Planck spectrum for $\rho(\om)$. Expressed in terms of $\rho(\om)$, Eq. (\ref{ode}) is the equation obtained by Einstein \cite{ein1}).

The fluctuation-dissipation relation (Eq.~\eqref{fd}),  can be used in the Fokker--Planck equation for the velocity distribution function $f(v,t)$ to obtain the Maxwell--Boltzmann distribution. That is, the Fokker--Planck equation with the drift and diffusion terms satisfying Eq. (\ref{fd}),
\be
\frac{\pa f}{\pa t}=-\xi\frac{\pa}{\pa v}(vf)+\frac{\xi k_BT}{m}\frac{\pa^2f}{\pa v^2},
\ee
has the solution
\be
f(v,t)=\bkt{\frac{m}{2\pi k_BT}}^{1/2}e^{-mv^2/2k_BT}
\ee
for $\xi t\gg 1$. The condition that the momentum diffusion and the drag force together determine the blackbody spectrum, subject to the assumption $\la\frac{1}{2}mv^2\ra=\frac{1}{2}k_BT$, therefore implies the full Maxwell--Boltzmann distribution for the particle velocities \cite{milne}.

As noted above, we have assumed throughout that $\alpha_I(\om)>0$, consistent with the optical theorem for Rayleigh scattering. This assumption does not allow for stimulated emission. However, we can derive the momentum diffusion constant and the drag force for atoms, allowing for stimulated emission, by a similar analysis in which $\alpha_I(\om)$ is introduced not through the optical theorem but through the response of the field to the atoms. The same expression Eq. (\ref{nonreldrag}) then follows for the drag force, for example, on an atom. For a two-level atom, for example, the polarizability in the rotating-wave approximation is 
\be
\alpha(\om)=\frac{{{\mu}}^2}{3\hbar}\frac{p_1-p_2}{\om_0-\om+i\beta},
\ee
\be
\alpha_I(\om)=\frac{{{\mu}}^2}{3\hbar}\frac{\beta(p_1-p_2)}{(\om_0-\om)^2+\beta^2},
\ee
where $\om_0$ and $\mu$ are the transition frequency and transition dipole moment, respectively, $\beta$ is proportional to the homogeneous linewidth, and $p_1$ and $p_2$ are respectively the lower- and upper-level occupation probabilities. Then, from Eq. (\ref{nonreldrag}),
\bea
F_{\rm ind}&\cong&-\frac{4\pi\om_0v}{c^2}\big[\rho(\om_0)-\frac{\om_0}{3}\frac{d\rho(\om_0)}{d\om}\big]
\int_0^{\infty}d\om\alpha_I(\om)\cong -\frac{4\pi^2\mu^2\om_0v}{3\hbar c^2}(p_1-p_2)\big[\rho(\om_0)-\frac{\om_0}{3}\frac{d\rho(\om_0)}{d\om}\big]\nonumber\\
&=&-\frac{\hbar\om_0}{c^2}(p_1-p_2)B\big[\rho(\om_0)-\frac{\om_0}{3}\frac{d\rho(\om_0)}{d\om}\big]v,
\eea
where $B=4\pi^2\mu^2/3\hbar^2$ is the Einstein $B$ coefficient. This is equivalent to the expression for the drag force obtained by Einstein \cite{ein1}.

\section{Acknowledgment}
We are grateful to Bruce Shore for his many and varied contributions to atomic, molecular, and optical science. Peter Milonni is especially thankful for having been a friend of Bruce's for many years.

\appendix

\section{Statistical independence of $E_z$ and $\pa E_z/\pa x$}
\label{App:stind}
Consider first the following elementary example. Let $X$ and $Y$ be two classical Gaussian random processes such that $\la X\ra=\la Y\ra=0$ and $\la XY\ra=0$. Then
$X$ and $Y$ are statistically independent, i.e., the joint probability distribution $p_{XY}(X,Y)=p_X(X)p_Y(Y)$. The simple proof of this goes as follows. Consider the characteristic function
\be
C_{XY}(\xi,\eta)=\Big\la \exp{i(\xi X+\eta Y)}\Big\ra
\ee
whose Fourier transform is $p_{XY}(X,Y)$. Since $X$ and $Y$ are Gaussian, so are linear combinations of $X$ and $Y$. So $Z=\xi X+\eta Y$ is Gaussian and $\la Z\ra=0$. Then
\bea
\Big\la \exp{(iZ)}\big\ra&=&1-\frac{1}{2}\la Z^2\ra+\frac{1}{24}3{\big\la Z^2\big\ra}^2 - ... =\exp\big(-\la Z^2\ra/2\big)\nonumber\\
&=&\exp\big(-\la(\xi X+\eta Y)^2/2\ra\big)=\exp\big(-\la \xi^2 X^2+\eta^2 Y^2+2\xi\eta XY\ra/2\big)\nonumber\\
&=&\exp\big(-\la \xi^2 X^2\ra/2\big)\exp(-\la \eta^2 Y^2\ra/2\big),
\label{eqa1}
\eea
since $\Big\la XY\Big\ra=0$ by assumption. We have used the fact that for Gaussian statistics $\la Z^n\ra=0$ for $n$ an odd positive integer and
\be
\Big\la Z^{2n}\Big\ra=1\cdot 3 \ ... \cdot (2n-1)\Big\la Z^2\Big\ra^n.
\ee
to write $\Big\la Z^4\Big\ra=3\Big\la Z^2\Big\ra^2$ etc. in (\ref{eqa1}). Thus $C_{XY}(\xi,\eta)=C_X(\xi)C_Y(\eta)$ and therefore $P_{XY}(X,Y)=p_X(X)p_Y(Y)$.

In the case of blackbody radiation, treated classically or quantum mechanically, the moments $\big\la E_z^n\big\ra$ have the Gaussian form as above. Since $E_z$ is Gaussian, so is its derivative $\pa E_z/\pa x$ according to a well-known property of Gaussian random processes. Furthermore 
$\big\la E_z(\pa E_z/\pa x)\big\ra=0$. Therefore, in accordance with the example above, $E_z$ and $\pa E_z/\pa x$ for blackbody radiation can be regarded as statistically independent in calculating expectation values, as shown by Einstein and Hopf \cite{eh1}.

\section{Momentum diffusion and decoherence from scattering by air molecules}
\label{App:air}

According to Campbell's theorem the momentum variance of dielectric particles due to momentum kicks from air molecules is
\eqn{d\bkt{\Delta p^2}\approx\dd q  (q^2)  \rho\bkt{q} \frac{q}{m_\mr{air}}\sigma_\mr{air}
,}
where $q = m_\mr{air} v$ is the momentum of an air molecule, $ \sigma_\mr{air} = 2\pi a^2 /3$ is the scattering cross section, and
\eqn{\rho(q) = \frac{N_\mr{air}}{V} 4\pi q^2 \bkt{\frac{1}{2\pi m_\mr{air} k_B T}}^{3/2} e^{-q^2/(2m_\mr{air}k_B T)} }
corresponds to the Maxwell-Boltzmann distribution for the air molecules.
Integrating over the velocities and using similar arguments as in the case of blackbody radiation, we obtain
\eqn{
\frac{\avg{\Delta p^2}}{\Delta t} \approx &\frac{8\pi}{m_\mr{air}}\frac{N_\mr{air}}{V} \bkt{\frac{1}{2\pi m_\mr{air} k_B T}}^{3/2}\int_0^\infty \dd q q^5 e^{-q^2/(2m_\mr{air}k_B T)}\\ 
= &\frac{16 a^2}{3}\frac{N_\mr{air}}{V} \sqrt{2\pi m_\mr{air }} \bkt{k_B T}^{3/2},}
which is twice the center-of-mass decoherence rate for scattering of air molecules in the long-wavelength limit  \cite{schloss}, similar to the case  of thermal photon scattering.


\begin{thebibliography}{99}
\bibitem{scalar} It is assumed that $\alpha(\om)$ is a scalar, so that the linear response to the electric field is independent of the field's direction.
\bibitem{notation} $e_{\bk\lambda z}$ is the $z$ component of the polarization unit vector ${\bf e}\rmk$ for a field mode with wave vector ${\bf k}$ and
angular frequency $\om =|{\bf k}|c$. ${\bf e}\rmk$ is taken to be real.
\end{thebibliography}
\end{document}